\documentclass[9pt,twocolumn,twoside]{osajnl}
\journal{josaa}

\setboolean{shortarticle}{false}

\usepackage[english]{babel}

\usepackage{epsfig}
\usepackage{nicefrac}
\usepackage[normalem]{ulem}

\usepackage{color}
\newcommand{\upd}[1]{{#1}}
\newcommand{\updb}[1]{{#1}}
\renewcommand{\doi}[1]{}

\def \uf{$\mbox{ m}^{-1}$}

\title{Useful relations for the analysis of stellar scintillation at the entrance pupil of a telescope}

\author[1,2]{Victor Kornilov}
\author[1]{Boris Safonov}
\author[1,3,*]{Matwey Kornilov}

\affil[1]{Lomonosov Moscow State University, Sternberg Astronomical Institute, Universitetsky prosp. 13, Moscow 119992, Russia}
\affil[2]{Lomonosov Moscow State University, Faculty of Physics, Leninskie Gory, 1-2, Moscow 119991, Russia}
\affil[3]{National Research University Higher School of Economics, \nicefrac{21}{4} Staraya Basmannaya, Moscow 105066, Russia}
\affil[*]{Corresponding author: matwey@sai.msu.ru}

\begin{abstract}

\upd{The development of new techniques for characterizing atmospheric optical turbulence (OT) has become an active topic of research again in recent years. In order to facilitate these studies, we reconsidered known theoretical results and \updb{obtained} some new practically useful \updb{conclusions}.
We introduce a dimensionless Fresnel filter, which allows us to approximate a polychromatic weighting function (WF) by a monochromatic one with a typical precision of several percent.}
\upd{A so-called dimensionless WF} can be easily scaled for a receiving aperture of any size. For the case of a circular aperture and monochromatic radiation, an analytical expression for the WF was found. The WFs for a square aperture and for a circular aperture match with relative  difference less than 0.01 if the circular aperture diameter is 1.15 times larger than the square aperture side.

\upd{A linear digital filter can be applied to the scintillation signal from an image detector.
As an example of digital filtering, we considered the power law filter $\propto f^{5/3}$ with the WF being constant in a wide range of altitudes. We discuss the main limitations of this approach for measuring OT integral: finite pixel size, aliasing, and finite image detector size.}

\end{abstract}

\setboolean{displaycopyright}{false}
 
\begin{document} 

\maketitle


\section{Introduction}
\label{sec:intro}

The phenomenon of atmospheric scintillation has been known since prehistoric times. It is especially apparent when the flux variations of stars (which are ideal point--like sources) are observed by naked eye. Light wave from such a source undergoes perturbations by optical inhomogeneities of the Earth's atmosphere, that leads to phase and amplitude fluctuations of the initially plane wavefront.

The generally accepted theory of atmospheric scintillation has existed since about the middle of the last century \cite{Tatarsky1967,Roddier1981}. The interest in further studies of its properties has not disappeared for many reasons. As a rule, the obtained results are of a general nature, but we will focus on stellar scintillation.

In the field of astronomy the study of scintillation pursuits two major goals. The first is to minimize an additional uncertainty of the flux measurements, so-called scintillation noise (e.g. \cite{Young1969,Dravins1997a,2012AA}). The second, a functional dependence of the amplitude variations on the light propagation distance makes stellar scintillation a powerful instrument for measuring the optical turbulence (OT) structure in the Earth's atmosphere. Additionally, the amplitude measurements are easier than phase ones. Despite the variety of tools developed for such measurements, they can be divided into two main groups.

\updb{Instruments of the first group use a two--dimensional detector, each element of which corresponds to some small subaperture at the telescope entrance pupil. Scintillation detection and ranging (SCIDAR) approach assumes use of a binary source \cite{Fuchs1998,Avila2008}. For more recent instruments, e.g. Full Aperture Scintillation Sensor (FASS) \cite{Guesalaga2020}, Ring-Image Next Generation Turbulence Sensor (RINGSS) \cite{Tokovinin2020},  Shack--Hartmann Multiaperture Scintillation Sensor (SH--MASS) \cite{Ogane2021} a single star is sufficient. } The instruments of the second group, which includes Multiaperture Scintillation Sensor (MASS) \cite{2003SPIE}, the sensor of lunar scintillation LuSci \cite{Rajagopal2008}, and others \cite{Ochs1976}, register the flux temporal fluctuations of the light passing through a set of apertures of certain shape and size.

The quantities measured by these instruments are the \updb{second--order} statistical moments of \updb{intensity fluctuations}, and usually it is required to obtain the OT characteristics at a position on the line of sight from these moments. Scintillation power is related to the intensity of the OT in a thin layer at the distance $z$ with help of so--called weighting function~(WF). For the real atmosphere, the inverse transformation from scintillation to the OT power is usually an ill-posed problem.

In this paper, we address practical issues associated with the scintillation measurement that are poorly described in the existing literature on this topic. First, we return to the effect of the detector response spectral width. Then, we consider the different ways of scaling the WF and present a dimensionless WF in analytical form. Next, we compare the WFs for square and circular apertures and show their interchangeability. \updb{Our work is limited only to variances of intensity fluctuations, the covariances are not considered.}

Section~\ref{sec:cdigital-filters} discusses the digital aperture filtering technique, \upd{relevant for measurement of scintillation with two--dimensional imaging arrays}. It provides an example of a digital filter that \upd{may produce} the WF constant over the entire range of propagation distances. \upd{The limitations of this approach arising due to properties of realistic detectors are investigated in Section~\ref{sec:wfs-digit}}.

\section{Dimensionless polychromatic Fresnel filter}
\label{sec:theory}

Under the assumption of weak perturbations, the scintillation index $s^2 = \left\langle (I-\langle I \rangle)^2\right\rangle/\langle I \rangle^2$ (the normalized variance of the fluctuations of the flux $I$) is uniquely defined by the distribution of structure coefficient of refractive index $C_n^2$ along the line of sight, e.g., \cite{Roddier1981}:
\begin{equation}
s^2 = \int_A C_n^2(z)\,Q(z) \,{\rm d}z,
\label{eq:s2int}
\end{equation}
here $z$ is the distance towards the turbulent layer, which is equal to the altitude of that layer above the observatory in case of measurement at zenith. The WF linking the scintillation effect at the surface with the OT contained in this layer is denoted by $Q(z)$. Integration is performed over the whole atmosphere. Note, that the case of unsaturated scintillation $s^2 \lesssim 0.1$ is assumed in equation~(\ref{eq:s2int}).

The technique for calculating the $Q(z)$ was described in detail previously, see, e.g.   \cite{Tokovinin2002b, Tokovinin2003} and is actively used in the MASS method \cite{2003MNRAS,2003SPIE,2007bMNRAS}. For the Kolmogorov turbulence, which is isotropic and locally homogeneous, and has two--dimensional phase power spectrum $\sim |{\bf f}|^{-11/3}$, the WF can be expressed as an integral in polar coordinates:
\begin{equation}
Q(z) = 9.61\int_0^\infty f^{-8/3}\,S(z,f) A(f) \,{\rm d}f.
\label{eq:qzint}
\end{equation}
Here $f = |{\bf f}|$ is the spatial frequency vector modulus, we adopt \uf\ for the units of this value. The aperture filter $A(f)$ (AF hereinafter) accounts for the spatial averaging introduced by the receiver aperture. \updb{In equation (\ref{eq:qzint}) averaging of the two--dimensional AF over the polar angle was performed. The averaged AF coincides with the initial AF in case of circularly symmetric apertures.}

\upd{
For instance, the AF of the circular aperture of diameter $D$ is the Airy function. For an annular aperture with central obscuration $\epsilon$ (i.e. with the inner diameter of $\epsilon D$) it is:
\begin{equation}
A(f) = (1-\epsilon^2)^{-2} \left[\frac{2\,J_1(\pi Df)}{\pi Df} - \epsilon^2 \frac{2\,J_1(\pi \epsilon Df)}{\pi\epsilon Df}\right]^2.
\label{eq:ary_ann}
\end{equation}}

Fresnel filter $S(z,f)$ describes the evolution of the wave amplitude fluctuations during the light propagation. For \upd{the spectral energy distribution} $F(\lambda)$ it is \cite{Tokovinin2003}:
\begin{equation}
S(z,f) = \left(\int_{F} \frac{F(\lambda)}{\lambda}\sin(\pi\lambda z f^2) \,{\rm d}\lambda\right)^2\! = \left(\mathcal F_{\mathrm{Im}}\left(\frac{zf^2}{2}\right)\right)^2.
\label{eq:snormal}
\end{equation}
Here $\mathcal F_{\mathrm{Im}}$ is an imaginary part of the Fourier transform of $F(\lambda)/\lambda$. In case of monochromatic radiation of wavelength $\lambda_0$ this filter reduces to well--known expression $S(z,f) = \sin^2 (\pi\lambda_0 z f^2)/\lambda_0^2$. The corresponding spatial scale equals to the primary Fresnel zone radius $r_\mathrm{F} = (\lambda_0 z)^{1/2}$.

When the detector has the wide spectral band, the simplest way is to integrate equation (\ref{eq:qzint}) directly. However analytical solutions --- even in form of asymptotic approximation --- facilitate the consideration of border cases and improve the understanding of the phenomenon as a whole.

We modify Fresnel filter (\ref{eq:snormal}) by introduction a dimensionless wavelength $\mu = \lambda/\lambda_0$, where $\lambda_0$ is some characteristic wavelength for a certain photometric band, the meaning of $\lambda_0$ will be explained \upd{below in~(\ref{eq:norm})}. After the substitution we get:
\upd{
\begin{equation}
S(z f^2) = \lambda_0^{-2}\left[\int \frac{\varPhi(\mu)}{\mu}\sin(\pi \mu \lambda_0\, z\, f^2) \,{\rm d}\mu\right]^2\!.
\label{eq:snormal1}
\end{equation}
}
Normalization of the $\varPhi(\mu)$ band is defined by the condition: 
\begin{equation}
\int F(\lambda)\,{\rm d}\lambda = \int \varPhi(\mu)\,{\rm d}\mu \equiv 1,
\label{eq:snormal2}
\end{equation} 
which leads to $\varPhi(\mu) =  \lambda_0 F(\lambda)$ and the factor $\lambda_0^{-2}$ before the integral.
After the usual variable substitution $u = (\lambda_0 z)^{1/2}f = r_\mathrm{F}f$, Fresnel filter becomes\upd{
\begin{equation}
S\left(\frac{u^2}{\lambda_0}\right) = \lambda_0^{-2}\left[\int \frac{\varPhi(\mu)}{\mu}\sin(\pi \mu u^2) \,{\rm d}\mu\right]^2\! = \lambda_0^{-2}\,E(u,\lambda_0).
\label{eq:snormal3}
\end{equation}}
We interpret the integral \upd{$E(u,\lambda_0)$} as a dimensionless Fresnel filter depending on $u$ only, \upd{we will show further how to fix the value for $\lambda_0$}.
In monochromatic case \upd{the filter} equals to $\sin^2(\pi u^2)$.
After the substitution of (\ref{eq:snormal3}) into the expression for $Q(z)$ we get:
\begin{equation}
Q(z) = 9.61\lambda_0{\!\!\strut}^{-7/6}\,z{\strut}^{5/6} \int_0^\infty \!u^{-8/3}\,A(u/r_\mathrm{F}) E(u, \lambda_0) \,{\rm d}u.
\label{eq:snormal4}
\end{equation}


In case of infinitely small aperture we can assume $A(u)\equiv 1$ and obtain:
\begin{equation}
Q_0(z) = 9.61\,\lambda_0{\!\strut}^{-7/6}\,z{\strut}^{5/6} \int_0^\infty u^{-8/3}\,E(u, \lambda_0)\,{\rm d}u.
\label{eq:i0norm}
\end{equation}
The integral on the right-hand side does not formally depend on any \upd{free} parameter, but in fact its value is related to the energy distribution of the detected radiation. We assume that its value is equal to the value for the monochromatic case with a certain {\it equivalent} wavelength $\lambda_0$:
\upd{
\begin{equation}
\widehat{Q}_0(z) = 19.21 \lambda_0{\!\strut}^{-7/6} z\strut^{5/6} = 9.61\frac{3}{10}\frac{\pi^{4/3}\Gamma(\frac{7}{12})}{\Gamma(\frac{11}{12})}\lambda_0{\!\strut}^{-7/6} z\strut^{5/6}.
\label{eq:lim0}
\end{equation}
We revealed the structure of the well--known coefficient $19.21$ in equation (\ref{eq:lim0}) to make further transformations more clear.}
\upd{Making asymptotic~(\ref{eq:lim0}) equal with WF~(\ref{eq:i0norm}) computed by direct integration, $\lambda_0$ is determined numerically by solving the following algebraic equation:
\begin{equation}
\int_0^\infty u^{-8/3}\,E(u, \lambda_0)\,{\rm d}u = \frac{3}{10}\frac{\pi^{4/3}\Gamma(\frac{7}{12})}{\Gamma(\frac{11}{12})}.
\label{eq:norm}
\end{equation}
}

\upd{
The values of $\lambda_0$ were calculated for three spectral bands best suited for scintillation measurements ($V$, $R$, and typical front-illuminated CCD without any filter). Stellar sources of the different spectral classes were used. The corresponding equivalent $\lambda_0$ values are given in Table~\ref{tab:eff_lams}.}


\upd{
One can note that the values $\lambda_0$ are slightly larger than the corresponding effective (in ordinary photometric meaning \cite{Straizys1992}) wavelengths $\lambda_\mathrm{eff}$ in all cases. Therefore, the resulting coefficient before $z^{5/6}$ is somewhat smaller.
However, the relative differences are small and amount to \updb{$0.05 \relbar 0.06$} in the case of $V$ filter.
These differences are largest for the case of no--filter operation, but even then, they are less than \updb{$0.11 \relbar 0.12$} for all possible spectral classes, see Table~\ref{tab:eff_lams}.}

\upd{
Therefore, it is feasible to use $\lambda_\mathrm{eff}$ instead of the equivalent $\lambda_0$ for many realistic cases. For the bands $V$ and $R$ one can use some fixed spectral energy distribution, as far as the relative difference between stars $B0\,V$ and $K5\,III$ is less than 0.03 and 0.07, respectively. However for the CCD detector without any filters, the relative difference reaches 0.36 which makes consideration of the object's color absolutely necessary.}

\section{Dimensionless weighting functions}
\label{sec:unitless}

\begin{table}[t!]
\caption{\updb{Effective wavelengths $\lambda_\mathrm{eff}$ and equivalent wavelengths $\lambda_0$ (in nm) for various spectral classes and three typical spectral bands implemented with the front illuminated CCD ICX424.\label{tab:eff_lams}} }
\medskip
\centering
\begin{tabular}{l|rrrrrr}
\hline\hline
Spectral class & \multicolumn{2}{c}{$V$} & \multicolumn{2}{c}{$R$} & \multicolumn{2}{c}{CCD}  \\
              & $\lambda_\mathrm{eff}$ & $\lambda_0$ & $\lambda_\mathrm{eff}$ & $\lambda_0$ & $\lambda_\mathrm{eff}$ & $\lambda_0$ \\
\hline 
Flat spectrum &  546 & $-$     &  743 & $-$     &  562 & $-$   \\
$B0\,V$       &  540 & 565     &  721 & 768     &  500 & 550   \\
$A0\,V$       &  542 & 567     &  727 & 776     &  524 & 578   \\
$F5\,V$       &  545 & 571     &  735 & 786     &  557 & 618   \\
$G5\,III$     &  548 & 574     &  740 & 794     &  590 & 656   \\
$K0\,III$     &  549 & 576     &  743 & 797     &  600 & 667   \\
$K5\,III$     &  553 & 580     &  755 & 812     &  642 & 714   \\
\hline\hline
\end{tabular}
\end{table}

\upd{
For the subsequent analysis it is convenient to introduce the dimensionless weighting function:
\begin{equation}
\label{eq:dimlessW}
W(r_{\mathrm{F}}/D) = Q(z)/{Q}_0(z).
\end{equation}
As long as the aperture size and the spatial frequency always appear in AFs as the product $Df$ (for instance, see Eq.~(\ref{eq:ary_ann}) and~(\ref{eq:squ_ann})), $A(u)$ depends on the ratio $D/r_\mathrm{F}$ after the transition to the dimensionless frequency $u = r_{\mathrm{F}} f$, so $D/r_\mathrm{F}$ is the only parameter which is responsible both for altitude dependence and aperture-size dependence of dimensionless WFs.
Hence, the same WF describes all apertures with a similar shape.
}

\begin{figure}[t!]
\centering
\psfig{figure=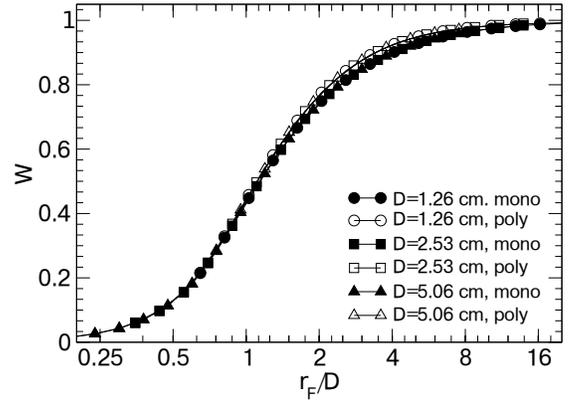,width=7.3cm}
\caption{\upd{Dimensionless WFs for $A0V$ star, $V$ band, as functions of $r_\mathrm{F}/D$ for various circular apertures (see diameters in the legend). Filled and empty symbols indicate monochromatic and polychromatic WFs, respectively, computed by direct integration. Analytical dimensionless WFs computed using equation~(\ref{eq:formulae}) are indistinguishable from the monochromatic curves in the scale of this graph.}
\label{fig:wfs_rat_a}}
\end{figure}

\upd{The examples of dimensionless WFs computed by numerical integration are presented in Fig.~\ref{fig:wfs_rat_a} as functions of $r_\mathrm{F}/D$. These WFs are calculated for the circular apertures of different sizes and a star of spectral type $A0\,V$ observed in $V$ band. For normalization we used $\lambda_0 = 567$~nm (see Table~\ref{tab:eff_lams}).
From the Figure, one can see how the functions $W(z)$ pass from one limit case to another along with the increase of the propagation distance.}



\upd{
Fig.~\ref{fig:wfs_rat_a} allows us to judge when the entrance aperture can be considered small and therefore the asymptotic WF for infinitely small aperture can be accurate enough. For example, when using Eq.~(\ref{eq:lim0}) instead of the exact dependence in the case of $r_\mathrm{F}/D = 3$ the maximum error is $\sim$0.15, for $r_\mathrm{F}/D = 5$ it decreases to 0.06. For the acceptable error of $\sim$0.10, the aperture with size of 1.1~cm can be considered small only for the altitudes higher than 3~km.
In case the square aperture of 3~cm (e.g. employed by \cite{Avila1997,Avila1998}) it is possible to have 0.1 uncertainty in the estimation of the OT power up to $z \sim 10$~km, in other words, this aperture cannot be considered to be small apriori.}

\upd{
In the case of circular aperture and monochromatic radiation the dimensionless WF can be expressed analytically with the aid of symbolic algebra software as follows:
\begin{multline}
W(\chi) = {}_4F_5\left( \left[-{\frac {5}{12}},\frac{1}{12},\frac{3}{4},\frac{5}{4}\right],
\left[\frac{1}{2},1,\frac{3}{2},\frac{3}{2},2\right],-\chi^2 \right) \\
-\frac {9216}{51425}\,\frac{\sin^3\left(\frac{5\pi}{12}\right)\sec\left(\frac{\pi}{12}\right) \csc\left(\frac{\pi}{12}\right) 2^{7/6} \left( \Gamma  \left( \frac {7}{12}\right)\right)^{2}\sqrt{3}}{\sqrt{\pi}\,\Gamma\left(\frac{2}{3}\right) 
\left(\Gamma\left(\frac {11}{12}\right)\right)^{2} }\chi^{5/6} \\
+\frac{5}{12}
\csc\left(\frac{\pi}{12}\right)\sin\left(\frac{5\pi}{12}\right)\,\chi 
\\\times\,{}_4F_5 \left(\left[\frac{1}{12},\frac{7}{12},\frac{5}{4},\frac{7}{4}\right], \left[\frac{3}{2},\frac{3}{2},2,2,\frac{5}{2}\right],-\chi^2 \right) =
\\ = {}_4F_5\left( \left[-{\frac {5}{12}},\frac{1}{12},\frac{3}{4},\frac{5}{4}\right],
\left[\frac{1}{2},1,\frac{3}{2},\frac{3}{2},2\right],-\chi^2 \right) \\ -2.195\,\chi^{5/6} \\
+1.555\,\chi 
\cdot\,{}_4F_5 \left(\left[\frac{1}{12},\frac{7}{12},\frac{5}{4},\frac{7}{4}\right], \left[\frac{3}{2},\frac{3}{2},2,2,\frac{5}{2}\right],-\chi^2 \right).
\label{eq:formulae}
\end{multline}
Here $\chi = \pi D^2/4\,r_\mathrm{F}^2$. The factor 4 in this formula appears due to the definition of $D$ as the diameter of the averaging aperture and the definition of Fresnel radius as a radius. Such notations arose historically. The comparison of the aperture radius and Fresnel radius would be more consistent: the factor 4 would disappear. The use of hypergeometric function ${}_4F_5$ should not be considered as a limitation, as long as much standard mathematical software usually support this calculation.
The curve computed using equation~(\ref{eq:formulae}) coincides with the numerically integrated monochromatic $W$ with precision of $2\times10^{-4}$.}

\begin{figure}[t!]
\centering
\psfig{figure=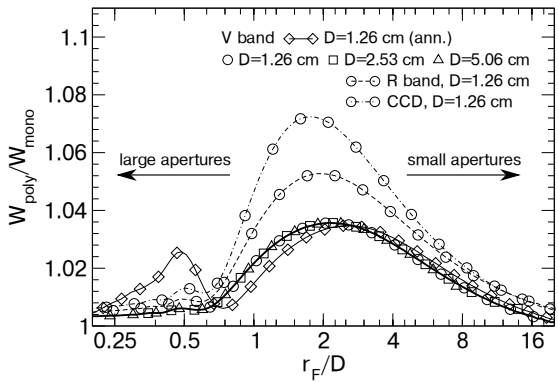,width=7.3cm}
\caption{\upd{Ratio of polychromatic dimensionless WF to monochromatic ones for various circumstances. Lines and symbols are explained in the legend.}
\label{fig:wfs_rat_b}}
\end{figure}

\upd{
Although dimensionless WFs for different $D$ are identical when expressed as functions of $r_\mathrm{F}/D$, there is a significant difference between WFs for polychromatic and monochromatic case, as one can note from Fig.~\ref{fig:wfs_rat_a}. This difference can be seen more clearly in Fig.~\ref{fig:wfs_rat_b}, where we plotted the ratio $W^{\prime}=W_\mathrm{poly}/W_\mathrm{mono}$ for different combinations of spectral bands, aperture shapes, and stellar spectra.}

\upd{
In the limits of large and small apertures, $r_\mathrm{F}/D\ll1$ and  $r_\mathrm{F}/D\gg1$, respectively, $W^{\prime}$ tends to unity, which means that polychromatic WF is well described by monochromatic one. It is well known that scintillation in a large aperture is almost achromatic. Besides, the application of such apertures is more typical for photometric studies than for OT measurements.}

\upd{
The maximum deviation is reached at intermediate values of normalized coordinate $r_\mathrm{F}/D\approx2$. For the basic case of $A0V$ star observed in $V$ band through a circular aperture, $W^{\prime}$ reaches the maximum of $1.04$. $W^{\prime}$ does not depend on aperture diameter (as expected).} 

\upd{
The $W^{\prime}$ dependence on $r_\mathrm{F}/D$ for other spectral bands has the same characteristic shape, but the amplitude is larger if the band is wider. The deviation reaches 1.05 and 1.07 for $R$ band and CCD without filter, respectively. Interestingly, the bimodal shape of $W^{\prime}$ becomes more prominent for an annular aperture.}

\begin{figure}[t!]
\centering
\psfig{figure=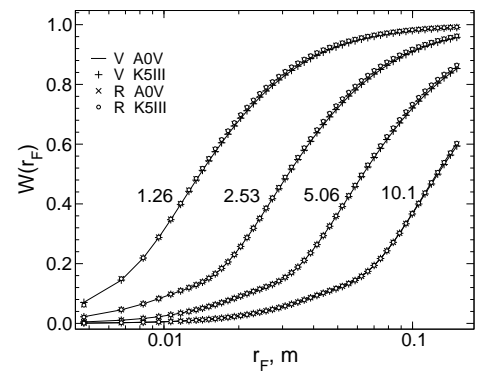,angle=-90,width=6.5cm} 
\caption{ Dimensionless WFs $W(r_\mathrm{F})$ for the circular aperture $D=1.26$~cm and three annular apertures with outer diameter of 2.53, 5.06, 10.1~cm and $\epsilon=0.5$. Evaluations is made for four energy distribution cases: white star $A0\,V$ and red star $K5\,III$, each in two bands $V$ and $R$. 
\label{fig:wfs_col_rf}}
\end{figure}

\upd{
The dependence of WF scaling on a spectral energy distribution is demonstrated in Fig.~\ref{fig:wfs_col_rf} using spectral classes $A0\,V$ and $K5\,III$ and photometric bands $V$ and $R$. This Figure shows the dimensionless WFs calculated for the circular aperture $D=1.26$~cm and three annular apertures $D=2.53, 5.06, 10.1$~cm with central obscuration of $\epsilon=0.5$.}

\upd{
The difference between white and red stars in the same photometric band does not exceed $2\cdot 10^{-3}$, which allows scaling the WFs using the single precomputed function $W(r_\mathrm{F})$, e.g. for the spectrum $A0\,V$ with the appropriate $\lambda_0$.}

\upd{
We conclude that the polychromatic WF can be approximated by the analytical monochromatic WF (Eq.~(\ref{eq:formulae})) with reasonable accuracy.}

\section{Aperture filters of circular and square apertures}
\label{sec:apertfun}

The spatial filters $A(f)$ are well--known for many aperture (entrance pupil) shapes, because they are identical to the diffraction image built in a focal plane by the aperture. The effect of the receiving aperture shape was discussed many times in the context of stellar scintillation studies, see e.g. \cite{Dravins1998} and detailed analysis in \cite{Tokovinin2002b}. 

\upd{The AF of a circular aperture is given by equation~(\ref{eq:ary_ann}).}
The spatial filter for a square aperture does not possess central symmetry, however it can be naturally expressed in Cartesian coordinates as $\mathrm{sinc}^2(af_x)\times \mathrm{sinc}^2(af_y)$. Here $a$ is the square side, and $\mathrm{sinc}(x) = \sin( \pi x)/(\pi x)$. In order to use this filter in equation (\ref{eq:qzint}) it is necessary to average the filter over the polar angle $\phi$.
\begin{equation}
A(f) = \frac{1}{2\pi}\int_0^{2\pi}\left[\mathrm{sinc}(af\cos \phi)\mathrm{sinc}(af\sin \phi)\right]^2\mathrm{d}\phi.
\label{eq:squ_ann}
\end{equation}

\begin{figure}[t!]
\centering
\psfig{figure=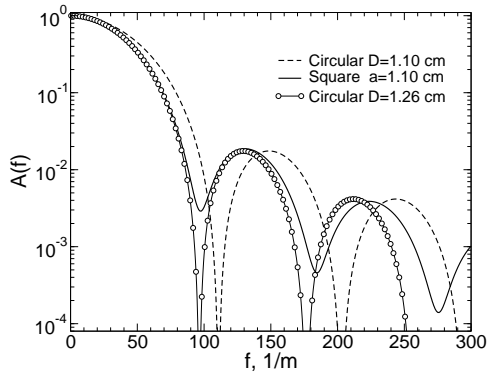,angle=-90,width=6.5cm} 
\caption{Aperture filters $A(f)$ for the circular apertures with $D=1.10$~cm and $D=1.26$~cm and square one with side of $a = 1.10$~cm.
\label{fig:apfs}}
\end{figure}

The AFs for square aperture $a = 1.10$~cm and circular one $D = 1.10$~cm without central obscuration are presented in Fig.~\ref{fig:apfs}. One can see that the behavior of these filters is mostly similar, the major difference is scale along $f$, that leads to a narrower main peak of $A(f)$ for the square aperture. This is expected as long as the area of square is $4/\pi$ times larger than the inscribed circle area.

Another difference is in high--frequency domain, where the functions peaks have little different shapes. After averaging, the values initially corresponding to the function zero points cease to be zero, although they remain quite small (less than $3\cdot10^{-3}$). 

It is possible to estimate the auxiliary frequency scaling factor for the correction of the main peaks widths analyzing Taylor series of (\ref{eq:ary_ann}) and (\ref{eq:squ_ann}). However, we determined this factor empirically using numerical data. It turned out that the AFs of circular and square apertures can be aligned by scaling the frequency $0.87$ times for circular aperture ($1.15$ difference in size), see Figure~\ref{fig:apfs}.

\upd{
In order to study the possibility to replace the WFs for square apertures with the WFs for circular apertures with modified dimension, the set of square apertures with sides of $0.55, 1.1, 2.2$ and $4.4$~cm has been considered. This set of apertures represents a frequently employed configuration of CCD operating without binning and with $2\times 2$, $4\times 4$, and $8\times 8$ binning. Entrance subapertures frequently used in practice (see the references in Introduction) have dimensions exactly in this range.}

\begin{figure}[t!]
\centering
\psfig{figure=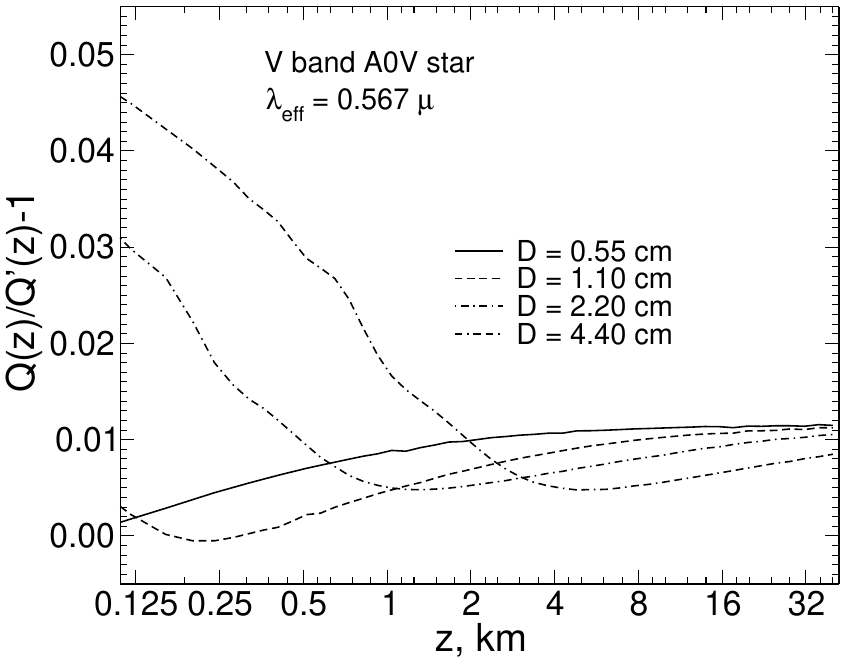,width=6.5cm} 
\caption{ Relative difference of the WFs $Q(z)$ for the square apertures and $Q^{\prime}(z)$ for the circular apertures of the test set, see the text. The calculation for $A0\,V$ star in photometric band $V$.
\label{fig:comps}}
\end{figure}

\upd{
Taking into account the coefficient $1.15$, these apertures corresponds to circular ones with diameters of $0.63$, $1.26$, $2.52$, and $5.06$~cm. The WFs were calculated for the case of $A0\,V$ star measured in $V$ band in order to compare them with each other, the results are presented in Fig.~\ref{fig:comps}}

\upd{
The comparison shows the systematic difference of $\sim 5\cdot 10^{-3}$ for the larger propagation distances $z$ since the curves were calculated with the different programs computing two-dimensional and one-dimensional integrals respectively. When $z$ becomes lower than $2$~km, there is a sharp rise of the relative difference because of the AFs difference in the high spatial frequencies domain. The difference is the most evident for the smallest apertures. For the aperture $1.1$~cm the WFs start to diverge at $z=0.5$~km and the relative difference becomes $\sim 1.045$ at $z=0$.}

\section{Digital aperture filters}
\label{sec:cdigital-filters}
 
Spatial filtering of scintillation is usually applied for the analysis of OT dependence on the altitude~\cite{Peskoff1968,Ochs1976,Tokovinin1998,2003SPIE}. As a rule, a diaphragm of the certain shape and size is installed in the pupil plane. It is possible to use several detectors corresponding to different apertures and combine their signal linearly. The resulting power can be computed using Eq.~(\ref{eq:qzint}) with the corresponding AF due to linearity.

In the case of the image detector with sufficient spatial resolution, installed in the telescope pupil, the scintillation signal processing can be effectively performed by means of Fourier transform. In the Fourier domain the aperture filtering reduces to the simple multiplication of the spectrum by the corresponding function. Such a filter does not need to be physically implementable in the pupil space.

A real implementation of digital filtering is naturally associated with additional limitations of the used detector and the atmospheric models. We will discuss this limitations in Section~\ref{sec:wfs-digit}. Meanwhile we demonstrate one purposely constructed AF. 

As before, we are using the stellar scintillation WF in the form of Eq.~(\ref{eq:qzint}). Whereas the $S(z,f)$ is given, we can affect the aperture filter $A(f)$ only. We would like to find the function $A(f)$ leading to the most smooth function $Q(z)$ as possible. Ideally, it is desirable to have $Q(z) = const$ for any $z \ne 0$, \upd{which would allow easy measuring of the total OT power on the line of sight without the systematic errors due to the non-uniform device response.}

Fortunately, $\tilde A(f) \equiv \mathrm{k} f^{5/3}$ leads to the required behavior, here $\mathrm{k}$ is some normalization constant, its dimension is $\mbox{m}^{5/3}$.  The fact can be proven strictly, but below we use a short explanation instead. Note, that Fresnel filter (\ref{eq:snormal}) has the symmetry property not only for monochromatic radiation but also for the polychromatic case~\cite{Tokovinin2003}.

It means that the following equation sequence is valid for any possible spectral input:
\begin{multline}
Q(z) = 9.61\int_{0}^{\infty} f^{-8/3} \tilde A(f) S(zf^2)\,{\rm d}f =\\
= \frac{9.61\mathrm{k}}{2} \int_{0}^{\infty} u^{-1} S(u) \,{\rm d}u = const,
\label{eq:int_from}
\end{multline}
where $u$ is substituted for $zf^2$ for $z \ne 0$. One may see that in the case under consideration the integral is independent of the distance $z$.

As far as the AF $\tilde A(f) = \mathrm{k} f^{5/3}$ indefinitely rises with frequency one may wonder if  integral (\ref{eq:int_from}) exists. Indeed, for monochromatic radiation $S(u)$ is a non-decreasing function of $u$ and this integral diverges at infinity. This divergence can be eliminated in two ways: 1) by limiting the integration limits 2) by introducing some damping factor in the Fresnel filter.

In the first case the dependence of the integral value on the frequency limit $g$ is
\begin{equation}
Q(z) = 2.40\,\mathrm{k}\,\lambda^{-2}\left(\gamma + \ln(2\pi\lambda z g^2) - \mathrm{Ci}(2\pi\lambda z g^2)\right),
\label{eq:int_cut}
\end{equation}
where $\gamma \approx 0.577$ is Euler constant, $\mathrm{Ci}$ is integral cosine.
We calculated the set of the WFs for values of $g$ from $10^{3}\mbox{ m}^{-1}$ to $10^{10}\mbox{ m}^{-1}$ using this equation. In the main range of altitudes, the WF rises logarithmically. At small values of $g$ the curves rise smoothly with some oscillations. For $g \gtrsim 10^{6}\mbox{ m}^{-1}$ the curves reach the region of smooth rise at very low altitudes (several meters). With further increase of $g$ they corresponds to ideal solution more and more. Typical value of the WFs for these curves is $\approx \mathrm{k}\cdot 10^{14}\mbox{ m}^{-1/3}$.

%
In order to test the second way we employed the polychromatic Fresnel filter $S(zf^2)=\lambda^{-2}\exp(-1.78z^2f^4\Lambda^2)\,\sin^2(\pi\lambda\,zf^2)$ from paper \cite{Tokovinin2003}. Here $\Lambda$ is the full width at half maximum of the spectral band. We get the following expression for the WF:
\begin{equation}
Q(z) = 13.3\mathrm{k}\,{}_2F_2\left([1,1],[\frac{3}{2},2], -\frac{\pi^2\lambda^2}{2\Lambda^2}\right)\Lambda^{-2}
\label{eq:int_polych}
\end{equation}
One can see that the WF is independent of the altitude and equals to $\mathrm{k}\cdot 0.3\cdot 10^{14}\mbox{ m}^{-1/3}$ for the case of the spectral band similar to $V$, which has $\lambda=570\mbox{ nm}, \Lambda=80\mbox{ nm}$.

Additional factors significantly affecting results may arise in real measurements of scintillation \upd{with image detectors. First,} it is the pixel structure of image sensors. Except for very exotic cases two--dimensional detector is the matrix of light-sensitive cells located in the nodes of a rectangular grid. These cells are usually square pixels, which form subapertures with size of $a\times a$ after the projection on the entrance pupil plane. Period of subapertures may differ from this size (e.g. when the detector with the interline transfer is used). However, it is usually assumed that the period equals to $a$ for both axes and defines the signal  sample spacing. \upd{The second factor is the finite area of recorded image.} The problems arising due to \upd{these facts} are discussed in \upd{the following Section.} 

\section{Specific details of digital spatial filtering}
\label{sec:wfs-digit}

\upd{
We consider CCD as a uniform two-dimensional array of pixels where each pixel is denoted by its column index $n$ and row index $m$.
In this case, the flux measurement will be an array of the intensities averaged over each pixel. The variance of the weighted fluctuations of these intensities can be considered as a synthetic scintillation index:
\begin{equation}
s=\Bigl\langle\Bigl(\sum_{n,m}\omega_{n,m}\Delta I_{n,m}\Bigr)^2\Bigr\rangle,
\end{equation}
where $\omega_{n,m}$ is a set of arbitrary real coefficients, $\Delta I_{n,m}$ is the relative flux variation for pixel $n,m$. 
}

\begin{figure}[t!]
\centering
\psfig{figure=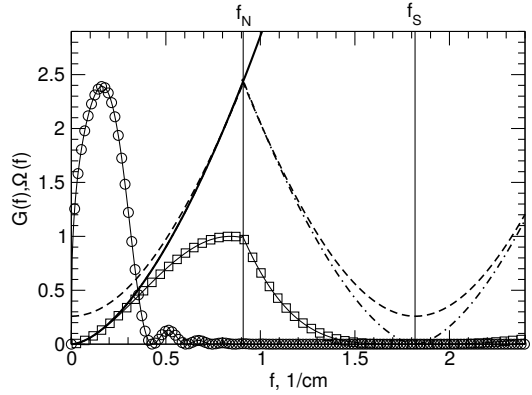,width=6.9cm}
\caption{\upd{
The section of scintillation power spectrum $G$ (\ref{eq:squarespec}) for square aperture with size $a=0.55$~cm along horizontal axis (line with circles). Calculations were made for $A0V$ star with $V$ band, propagation distance $z=1$~km. Thick solid line stands for $\mathrm{k}f^{5/3}$ filter introduced in Section~\ref{sec:cdigital-filters}. Dash--dotted line stands for approximation (\ref{eq:digitalFourier}) of this filter for the infinite array of pixels $a$, \updb{where $\omega_{n,m}$ were obtained using equation (\ref{eq:dnm_coefs})}. 
Solid line with squares is the same, multiplied by AF of square aperture $a$. Dashed line indicates $\Omega$ for array of $10$ pixels. Vertical lines indicate Nyquist frequency $1/(2a)$ and sampling frequency $1/a$. See the text for details.}
\label{fig:spectrum}}
\end{figure}

\upd{
In appendix \ref{app:dfilter} it is shown that this scintillation index depends on turbulence profile through the weighting function (\ref{eq:2dWF}) with the aperture filter of the following form: 
\begin{equation}
A(\mathbf{f}) = \left|\widetilde{P}(\mathbf{f}) \Omega(\mathbf{f})\right|^2.
\label{eq:digitalApertureFunction}
\end{equation}
Here the two--dimensional form of aperture filter is more adequate for the description of the CCD pixels array, which lacks the rotational symmetry. $\widetilde{P}(\mathbf{f})$ is the Fourier transform of aperture function of a single pixel, $\Omega(\mathbf{f})$ is the digital linear transform filter:
\begin{equation}
\Omega(\mathbf{f}) \equiv \sum_{n,m} \omega_{n,m} \exp\left[2\pi i a \left(f_x n + f_y m\right)\right],
\label{eq:digitalFourier}
\end{equation}
where $a$ is the distance between adjacent pixels. This representation can be used to simulate the performance of $\mathrm{k}f^{5/3}$ filter introduced in the previous section. The coefficients can be estimated, for instance, by inversion of Fourier transform, see equation~(\ref{eq:dnm_coefs}).}

\upd{
For the discussion of quantitative properties of filter $A(\boldsymbol{f})$ we need the normalization constant $\mathrm{k}$. The aperture function (\ref{eq:digitalApertureFunction}) reaches the maximum close to Nyquist frequency $f_\mathrm{N}=1/(2a)$, see the line with squares in Fig.~\ref{fig:spectrum}. It can be considered as a high--pass filter resembling differential scintillation filter \cite{Tokovinin2002b}.
In order to determine the maximum position, we differentiate the AF (\ref{eq:digitalApertureFunction}) over the frequency component $f_\mathrm{x}$, while $f_\mathrm{y}=0$. It is easy to show that the condition for the first maximum is:
\begin{equation}
\frac{\pi}{2}\frac{f_\mathrm{x}}{f_\mathrm{N}} = \frac{1}{6}\tan{\left(\frac{\pi}{2}\frac{f_\mathrm{x}}{f_\mathrm{N}}\right)},
\label{fig:derative}
\end{equation}
where the pixel size $a$ is replaced with the corresponding Nyquist frequency.}

This equation can be solved iteratively. The maximum of the AF is reached at frequency $f_\mathrm{max} = 0.927\,f_\mathrm{N}$. Substituting it into the AF yields:
\begin{equation}
A(f_\mathrm{max}) = \frac{4\cdot 1.0254}{\pi^2}\sin^2\left(\frac{\pi}{2}\,0.92745\right)\,f_\mathrm{N}^{5/3}.
\label{fig:normalization}
\end{equation}

Demanding that AF equal to unity at maximum, we get the value of the normalization constant $\mathrm{k} = 2.438\,f_\mathrm{N}^{-5/3}$. For the $0.55$~cm pixel, the constant is $\mathrm{k} = 0.0013\mbox{ m}^{5/3}$. Thus,  made in Section~\ref{sec:cdigital-filters} estimations $\sim 10^{11}\mbox{ m}^{-1/3}$ of the WFs values fall into the typical WFs amplitude range.

\upd{
The limitation of the expansion (\ref{eq:digitalFourier}) stems from two facts. First, $n$ and $m$ are integer numbers, therefore resulting $\Omega$ is a periodic function with the period known as sampling frequency $f_\mathrm{S}=1/a$, see Fig.~\ref{fig:spectrum}. Because of this $\Omega$ cannot reproduce the high--frequency behavior of $\mathrm{k}f^{5/3}$ filter. This effect is known as aliasing, which is typical for the analysis of sampled signals.}

\begin{figure}[t!]
\centering
\psfig{figure=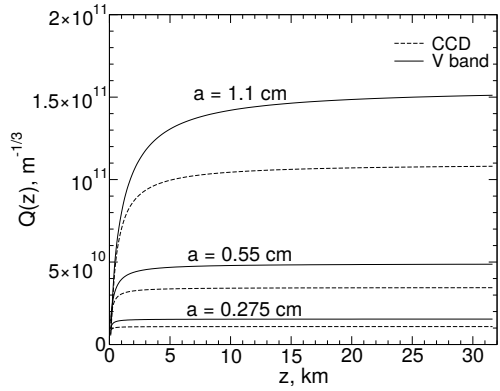,width=6.5cm}
\caption{Computed WFs for the scintillation signal after spatial filtering by AF $A_{\scriptscriptstyle\Box}({f_\mathrm{x}})\tilde A({f_\mathrm{x}})$. The pixel size is indicated on the plot. Calculations were made for $A0\,V$ star with $V$ band and with the CCD without any filter.
\label{fig:wf-final}}
\end{figure}

\updb{The scintillation power spectrum $G$ for square aperture is:
\begin{equation}
G(f) = 0.00961\,(2\pi)^2 f^{-11/3}\,S(z,f) |\widetilde{P}(\mathbf{f})|^2.
\label{eq:squarespec}
\end{equation}
It is plotted for propagation distance $z=1$~km and pixel size $a=0.55$~cm in Fig.~\ref{fig:spectrum}. One can see that the most of scintillation power is contained far below half the sampling frequency in this case.} Thus the effect of aliasing is expected to be low. We investigate it further by computing WFs for different apertures and spectral bands, see Fig.~\ref{fig:wf-final}.

As one could expect, in the main range of altitudes the WFs are constant at the level $Q_\infty$. In the region of small altitudes the WFs decreases due to impact of aperture filter of the pixel.  The level of $0.8\,Q_\infty$ is reached by the WF for $a=0.275$~cm at 100~m in full spectral band of the CCD detector. For subaperture $0.55$~cm the same is true for altitudes at 400~m for the CCD band and at 860~m for $V$ band. $Q_\infty$ values change from one curve to another by an order of magnitude, that allows us to chose the sensitivity to the scintillation adjusting the detector binning. Note that for these filtering the WF levels is smaller than for the differential scintillation measurements.

For scintillation generated by turbulence layers lower than 1~km the aliasing problem is more significant. Thus for altitudes of $\sim 250$~m it is preferable to have smaller pixels, $0.2 \relbar 0.3$~cm. Further decreasing of the aperture size may be undesirable because of the turbulence inner scale $l_0$. In this case the high--frequency behavior of the OT power spectrum is usually modeled by von Karman law with the~factor~ $\exp(-|{\bf f}|^2 l_0^2)$~(see,~e.g.~\cite{Masciadri1997AO}).

\begin{figure}[t!]
\centering
\psfig{figure=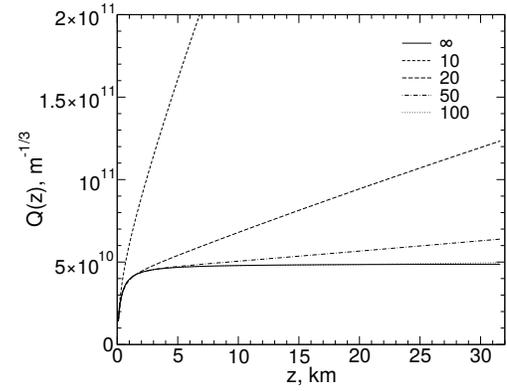,width=6.5cm}
\caption{\upd{The same as in Fig.~\ref{fig:wf-final}, but for different window size, which is indicated in legend in pixels. Calculations were made for $A0\,V$ star with $V$ band, pixel size is 0.55~cm.}
\label{fig:wf-window}}
\end{figure}

\upd{
The second issue with expansion (\ref{eq:digitalFourier}) is related to the finite number of pixels in the array $M\times M$ used for sampling of scintillation.
Because of that, we have to limit the expansion (\ref{eq:digitalFourier}): $n$ and $m$ run from $-M/2$ to $M/2$. The lower the $M$, the poorer the approximation of $\mathrm{k}f^{5/3}$ by $\Omega$, as one can see from Fig.~\ref{fig:spectrum}. This behavior is especially noticeable around $f=0$, where the original AF has a discontinuity of the first derivative. The WFs corresponding to AF for different array sizes are displayed in Fig.~\ref{fig:wf-window}. For small arrays, the WF rises linearly. However, starting with $M=50$ it can be considered constant.}

\upd{
We conclude that the $\mathrm{k}f^{5/3}$ filtering proposed in Section~\ref{sec:cdigital-filters} can be implemented with a two-dimensional array given that pixels are sufficiently small and the number of pixels is sufficiently large.
With a $50\times50$ array of 0.55~cm pixels the resulting WF changes by 10\% between 2 and 16 km.}

\section{Conclusions}
\label{sec:conclusions}

Let us summarize main features of scintillation measured through small apertures (subapertures). However, we note that these conclusions are correct for large apertures as well.

\begin{itemize}

\item \upd{A polychromatic WF can be approximated by a monochromatic one} with the equivalent $\lambda_0$ slightly different from the effective wavelength $\lambda_\mathrm{eff}$ of the spectral band. \upd{The precision of the approximation depends on the radiation spectral band and the aperture geometry.
In the case of circular filled aperture and observations with CCD and $V$ band the precision is 3.5\%.
We did not consider this approximation for differential scintillation indices used in MASS method, the effect of finite spectral bandwidth is expected to be much stronger in this case.} Even though the difference between $\lambda_0$ and $\lambda_\mathrm{eff}$ is non--zero, it can be neglected in \upd{some} cases.


\item The simple scaling allows one to transform a WF $Q(z)$ into dimensionless one $W(r_\mathrm{F}/D)$. For this, the WF $Q_0(z)$ for infinitely small entrance aperture should be used.
\upd{Using equation~(\ref{eq:dimlessW}),} one can easily obtain $Q(z)$ from the dimensionless WF $W(r_\mathrm{F}/D)$ for any geometrically similar apertures. For wavelength scaling in polychromatic case one would need the similarity of the spectral band as well.

\item For circular entrance aperture and monochromatic radiation, the analytic expression for $W(r_\mathrm{F}/D)$ is obtained in terms of the hypergeometric functions ${}_4F_5$. It was validated against the numerical integration.

\item The aperture filters (AF) and the WF for the square aperture with side of $a$ and the circular aperture with diameter of $D$ match very well if additional factor is employed: $D = 1.15\cdot a$. The relative difference is less than 0.01 in the most cases.

\item \upd{The WF for the aperture filter $\mathrm{k}f^{5/3}$ does not depend on altitude for any spectral band.
The $\mathrm{k}f^{5/3}$ filter can be applied for a reasonably accurate estimation of the OT integral over the line of sight, and subsequently seeing value.
Such a filter can be implemented by digitally processing the scintillation measurements carried with two-dimensional array of square subapertures, pixels.}

\item \upd{Additional averaging imposed by the pixels of the array limits the performance of $\mathrm{k}f^{5/3}$ filter:
the actual WF decreases at low altitudes.
This effect is less prominent for wider spectral bands.
As an example, for an array of 0.55~cm pixels and measurement with a CCD without any filter, the WF decreases by 10\% at 1.1~km.
The problem of suppression of WF at low altitudes may be alleviated by conjugation of the entrance pupil to the necessary negative altitude.}



\item \updb{Inner scale of the turbulence is another factor suppressing high--frequency OT, so decreasing subaperture size below $0.2 \relbar 0.3$~cm is useless. We note that the shrinking of the subapertures also leads to the rise of the impact of the detector noise and photon noise.}


\item \upd{Another factor limiting the implementation of scintillation digital filters is the finite number of pixels constituting the array.
We showed that the WF of $\mathrm{k}f^{5/3}$ filter rises linearly in the whole range of altitudes for small arrays.
For arrays larger than $50\times50$ the WF changes across the altitudes range under investigation is less than 10\%.}

\item \upd{
Although both finite size of pixel and finite number of pixels undoubtedly affect the performance of digital filters, we show that their effect can be quantified using the formalism of weighting functions. We do not exclude the possibility that there are approximations of $\mathrm{k}f^{5/3}$ by coefficients $\omega_{n,m}$, which perform better than equation (\ref{eq:digitalD}). These improved approximations may yield more constant behavior of WF or, alternatively, may be used for OT profile reconstruction, as in FASS~\cite{Guesalaga2020} or RINGSS~\cite{Tokovinin2020} methods.}

\end{itemize}

Registering the scintillation at the entrance pupil with subsequent digital processing becomes more and more widespread due to advantages of the modern image detectors. However one has to take into account that such technique is more prone to the detector noises than the classical approach employing an optical preprocessing of light and small number of detector channels. Thus the selection of detector for the instrument should be done with a great precaution.

A typical image detector does not fully exploit the basic properties of scintillation following from the OT homogeneity and isotropy. For example, a one--dimensional multi--element detector is sufficient for the estimation of the scintillation power spectrum. A CCD can emulate such detector increasing signal/noise ratio greatly, meanwhile a more trending CMOS detector cannot.

\appendix

\section{Digital filter response}
\label{app:dfilter}

\upd{
We consider CCD as a uniform two-dimensional array of pixels where each pixel is denoted by its column index $n$ and row index $m$.
Following~\cite{Tokovinin2003} we consider detected light flux in pixel $n,m$ normalized by its average,
and calculate the flux for weak scintillation $(\chi \ll 1)$:
\begin{equation}
I_{n,m} = \int \exp\left[2 \chi(\mathbf{r}) \right] P_{n,m}(\mathbf{r}) \,{\rm d^2} \mathbf{r} \approx 1 + 2 \bar{\chi}_{n,m},
\end{equation}
where
\begin{equation}
\bar{\chi}_{n,m} = \int \chi(\mathbf{r}) P_{n,m}(\mathbf{r}) \,{\rm d^2} \mathbf{r},
\end{equation}
$\chi(\mathbf{r})$ is the amplitude perturbation field,
$P_{n,m}(\mathbf{r})$ is the aperture function (AF) for pixel $n,m$.
All pixels have the same aperture shape, so
\begin{equation}
P_{n,m}(\mathbf{r}) = P(x - a n, y - a m),
\end{equation}
where $a$ is the pixel size, and the single pixel AF $P(\mathbf{r})$ is symmetric.
Employing the Fourier~transform~(FT) shift property we get the following:
\begin{equation}
\widetilde{P}_{n,m}(\mathbf{f}) = \widetilde{P}(\mathbf{f}) \exp\left[-2\pi i a \left(f_x n + f_y m\right)\right].
\end{equation}
Here $\widetilde{P}$ is FT of the AF, $\mathbf{f}$ is spatial frequency.
Then $\bar{\chi}_{n,m}$ is represented as the following
\begin{equation}
\bar{\chi}_{n,m} = \int \widetilde\phi(\mathbf{f}) \sin(\pi\lambda z \left|\mathbf{f}\right|^2) \widetilde{P}(\mathbf{f}) \exp\left[2\pi i a \left(f_x n + f_y m\right)\right] \,{\rm d^2}  \mathbf{f},
\end{equation}
where $\widetilde\phi(\mathbf{f})$ is the FT of phase perturbation field,
$\sin(\pi\lambda z \left|\mathbf{f}\right|^2)$ is Fresnel diffraction filter.}

\upd{
We introduce the digital scintillation index $s \equiv \left<\Delta I^2\right>$
where the digitally weighted flux variation $\Delta I$ is defined as follows
\begin{equation}
\Delta I \equiv \sum_{n,m} \omega_{n,m} \Delta I_{n,m} = \sum_{n,m} \omega_{n,m} \left(I_{n,m} - 1 \right),
\end{equation}
where $\omega_{n,m}$ is a set of arbitrary real coefficients, $\Delta I_{n,m}$ is the flux variation for pixel $n,m$.
Here the angle brackets denote ensemble averaging.}

\upd{
Note, that
\begin{equation}
\label{eq:delta_I}
\Delta I = 2 \int \widetilde\phi(\mathbf{f}) \sin(\pi\lambda z \left|\mathbf{f}\right|^2) \widetilde{P}(\mathbf{f}) \Omega(\mathbf{f}) \,{\rm d^2}  \mathbf{f},
\end{equation}
where the digital filter $\Omega(\mathbf{f})$ depends on the parameters $\omega_{n,m}$:
\begin{equation}
\label{eq:digitalD}
\Omega(\mathbf{f}) \equiv \sum_{n,m} \omega_{n,m} \exp\left[2\pi i a \left(f_x n + f_y m\right)\right].
\end{equation}}

\upd{
If the desired digital filter $\Omega(\mathbf{f})$ can be expressed in form~(\ref{eq:digitalD}),
then the parameters can be evaluated as follows:
\begin{equation}
\omega_{n,m} = a^2 \int_{-1/(2a)}^{1/(2a)} \Omega(\mathbf{f}) \exp\left[-2\pi i a \left(f_x n + f_y m\right)\right]  \,{\rm d^2} \mathbf{f}.
\label{eq:dnm_coefs}
\end{equation}
Otherwise, $\omega_{n,m}$ can be found as a solution of some function approximation problem.}

\upd{
Equation~(\ref{eq:delta_I}) is a counterpart of Eq.~(7) in~\cite{Tokovinin2003}.
Following the same approach as~\cite{Tokovinin2003} to account for polychromatic scintillation,
we find that
\begin{equation}
s = \int Q(z) C^2_n(z)  \,{\rm d}z,
\end{equation}
where
\begin{equation}
Q(z) = 0.00961\,(2\pi)^2 \int f^{-11/3}\,S(z,f) A(\mathbf{f}) \,{\rm d^2}\mathbf{f}.
\label{eq:2dWF}
\end{equation}
Here $f = \left|\mathbf{f}\right|$, $z$ is the altitude, and $A(\mathbf{f})$ is $\left|\widetilde{P}(\mathbf{f}) \Omega(\mathbf{f})\right|^2$.}

\begin{backmatter}
\bmsection{Acknowledgments} \updb{We are grateful to the referees for the comments which stimulated improving analysis and representation.}

\bmsection{Disclosures}
The authors declare no conflicts of interest.

\bmsection{Data availability}
No data were generated or analyzed in the presented research.
\end{backmatter}

\bibliography{My_paper_list,new-scint-enNotes,astroclimatic}

\end{document}